\documentclass[12pt]{article}
\usepackage{graphicx,psfrag,amssymb,amsmath,setspace}
\newcommand{\appropto}{\mathrel{\vcenter{
  \offinterlineskip\halign{\hfil$##$\cr
    \propto\cr\noalign{\kern2pt}\sim\cr\noalign{\kern-2pt}}}}}
\begin{document}
\psfrag{sigmav}{$\langle \sigma v \rangle$}
\psfrag{cm3s}{$\rm cm^3/s$}
\psfrag{Ratio}{$\langle \sigma v\rangle_{DT}/\langle \sigma v\rangle_{DD}$}
\title{Plasma Temperature Inference from DT/DD Neutron Discrimination}
\author{J. I. Katz\thanks{
Proofs to: J. I. Katz, Department of Physics, CB1105, Washington University,
1 Brookings Dr., St. Louis, Mo. 63130.  19 pp., 3 figures, 1 table
{\tt katz@wuphys.wustl.edu} facs: 314-935-6219}
\\ Los Alamos National Laboratory \\ P. O. Box 1663,
Los Alamos, N. Mex. 87545 \\ and \\ Department of Physics \\ and \\
McDonnell Center for the Space Sciences \\ Washington University, St. Louis,
Mo. 63130}
\maketitle
\newpage
Plasma Temperature Inference from DT/DD Neutron Discrimination

J. I. Katz

\begin{abstract}
DD and DT reaction rates may be compared to determine plasma temperatures in
the 10--200 eV range.  Distinguishing neutrons from these two reactions is
difficult when yields are low or unpredictable.  Time of flight methods
fail if the source is extended in time.  These neutrons may be distinguished
because inelastic scattering of more energetic neutrons by carbon produces
a 4.44 Mev gamma-ray, and because hydrogenous material preferentially
attenuates lower energy neutrons.  We describe a detector system that can
discriminate between lower and higher energy neutrons for fluences as low
as $\mathcal{O}(10^2)$ neutrons/sterad even when time of flight methods
fail, define a figure of merit and calculate its performance over a
broad range of parameters.
\end{abstract}
\newpage
\section{Introduction}
Plasma temperatures in the range 10--200 eV can be difficult to measure
because at those temperatures thermal emission is in the vacuum UV and soft
X-ray bands and is strongly absorbed by matter.  This problem is acute if
the plasma is surrounded by cold dense matter.  For example, opaque cold
dense metallic imploding liners may be used to compress and heat plasmas,
with the ultimate goal of thermonuclear fusion \cite{T08}.

We consider the problem of measuring plasma temperatures in the approximate
range 10---200 eV if radiation from the plasma is not directly observable.
At these temperatures the very soft X-rays of thermal emission are strongly
absorbed by most substances (including vacuum windows, for example), so that
this circumstance may be frequently encountered.  If the plasma contains a
suitable admixture of deuterium and tritium, comparison of the rate of the
D(D,$^3$He)n reaction, producing 2.45 MeV neutrons, to that of the
D(T,$\alpha$)n reaction, producing 14.1 MeV neutrons, may permit the
determination of the temperature.

The ratio of these reaction rates is significantly dependent on temperature.
The steep increase of both reaction rates with increasing temperature in
this regime implies that the temperature determined is that of the highest
temperature encountered in the system, even though that may be found in only
a small fraction of its volume.  It is also insensitive to other parameters
such as the volume and density at peak temperature and the confinement time.

Determining plasma temperature by comparison of DD and DT reactions in this
low temperature regime involves some special problems.  Because of the
extreme sensitivity of the reaction rates to temperature, the total number
of neutrons produced may be small.  In addition, it may not be predictable
even in order of magnitude because small uncertainties in the temperature
correspond to great uncertainties in the reaction rates.  The small number
of expected neutrons requires that detectors be close to the source, and
sources produced by, for example, comparatively massive imploding liners may
have long lifetimes.  For these reasons the usual method of discriminating
neutrons of different energy, their times of flight, may not be feasible.
After examining this issue, this paper discusses other methods.  
Discrimination based on the ability of the more energetic DT neutrons to
excite the 4.44 MeV state of $^{12}$C and on the greater elastic scattering
cross-section of lower energy neutrons in hydrogen may be the most feasible
method.  A detector design is outlined, a figure of merit defined, and
quantitative simulation results are presented.

\begin{figure}[h!]
\begin{center}
\includegraphics[width=4in]{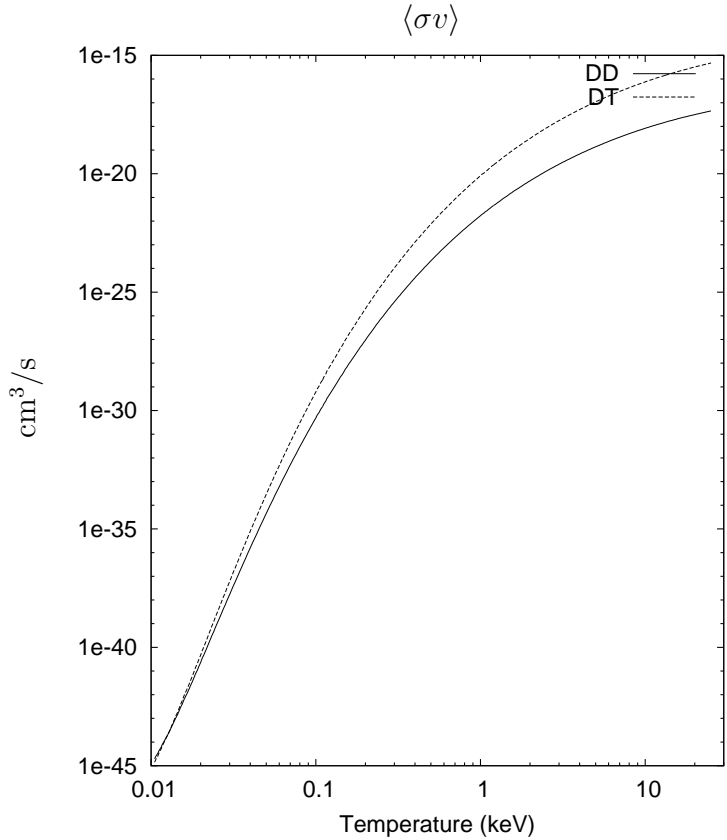}
\end{center}
\caption{\label{reactrate}Maxwell-averaged reaction rate coefficients
\cite{NRL07}, showing their steep dependence on temperature at low
temperature.  Electron shielding \cite{SVH69} in weak and intermediate
regimes is included for an assumed density of 100 gm/cm$^3$, as in some
inertial fusion targets.}
\end{figure}

Because this method depends on the thermonuclear reaction rates, these are
shown in Figure \ref{reactrate} for $k_B T < 25$ keV.  At these temperatures
the nonresonant expressions given by \cite{NRL07} are valid, and the Figure
extends to lower temperatures than are usually shown.  It is also necessary
to include the effects of electron screening \cite{SVH69}, which increases
the reaction rates at the lowest temperatures considered by several orders
of magnitude at high, but plausible, densities.

\begin{figure}[h!]
\begin{center}
\includegraphics[width=4in]{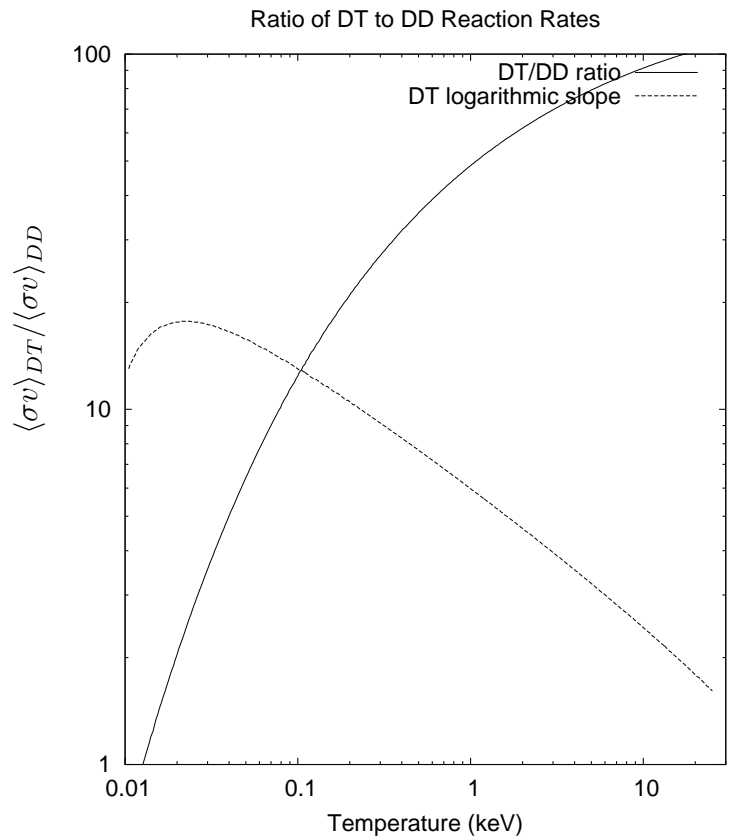}
\end{center}
\caption{\label{reactratio}Ratio of DT to DD reaction rates and logarithmic
derivative of the DT rate with respect to temperature.}
\end{figure}

Figure \ref{reactratio} shows the ratio of the DT to the DD reaction rates
and the logarithmic derivative $\partial\ln{\langle \sigma v \rangle} /
\partial \ln{T}$ of the DT reaction rate with respect to temperature (the
result for the DD rate is very similar).  For comparatively low ($\lesssim
200\,$eV) temperatures the ratio of the fluence of DT to DD neutrons
unambiguously determines the temperature in the reaction region, independent
of any other parameter, as shown in Fig.~\ref{reactratio}.
\section{Distinguishing DT from DD neutrons}
\subsection{Time of Flight}
At low temperatures the number of neutrons produced is small.  The usual
method of determining source temperatures from the width of the arrival time
distribution at a detector of neutrons produced by a single reaction is
difficult or impossible, partly because the weak source requires a detector
close to it, reducing the spread of arrival times, partly because low
temperatures reduce that spread further, and most importantly because
of the poor statistics when only a small number of neutrons are detected.

For example, at a distance of 1 m and a source temperature of 100 eV the
$1/e$ half-width of the arrival time distribution of DT neutrons is 23 ps,
while that of DD neutrons is 147 ps.  These widths are the flight times
multiplied by $\sqrt{2k_B T/M v_n^2} = \sqrt{k_B T /(E_n A)}$, where $M$ is
the total mass of the reactants, $v_n$ the neutron velocity, $E_n$ the
neutron energy and $A$ the sum of the atomic numbers of the reactants.  For
the DT reaction this is 0.0012, while for the DD reaction it is 0.0032; if
the source is distributed over a region whose size is this fraction of the
distance to the detector (1.2 mm and 3.2 mm, respectively, for DT and DD
reactions at 1 m distance) then the variation in path lengths will wash out
any information about the source temperature obtainable from time of flight.

The difference in time of flight between 2.45 MeV (DD) and 14.1 MeV (DT)
neutrons is much larger, and if both reactions occur might be used to
distinguish these two populations and to infer the temperature from the
ratio of their reaction rates.  At 1 m range the 14.1 MeV DT neutrons arrive
after 19 ns, while the 2.45 MeV DD neutrons arrive after 46 ns, a lag of 27
ns.  This interval is measurable with plastic scintillators whose response
times may be 2--10 ns.  However, separation by time of flight, even into
2.45 and 14.1 MeV energy groups, is possible only if the source duration is
less than the interval between the arrivals of 14.1 MeV and 2.45 MeV
neutrons.  If the source duration is comparable to or greater than this
interval, as in imploding liner experiments in which neutrons may be emitted
over a time $\gtrsim 100\,$ns \cite{T08}, another method of discrimination
is required.
\subsection{Energy Deposited}
For sources with long ($\gtrsim 30 d\,$ns) durations at a distance $d$ (in
m) discrimination of DT from DD neutrons is not possible by time of flight
alone, even for an ideal detector.  Discrimination on the basis of the
energy deposited could be possible, provided the detector were segmented
so that $\lesssim 1$ neutron interacts in each segment.  This would require
$\gtrsim N$ segments, where $N$ is the total number of detected neutrons.
If the fluences were approximately predictable, the segment sizes could be
chosen to keep $N$ in a feasible range.

To determine the ratio of DT to DD neutrons to a fractional accuracy $f$ at
$s$ standard deviations significance, for the optimal case in which this
ratio is $\approx 1$, requires
\begin{equation}
N \gtrsim {4 s^2 \over f^2}.
\end{equation}
For $f = 0.1$ (even though the reaction rates are very sensitive to 
temperature their ratio is not) and $s = 2$ we find $N > 1600$.  The number
of discrete segments must be comparable (the large ratio of energy between
the two neutron groups permits some discrimination even when two or three
neutrons are detected in a single segment).  

This might be feasible (if the plastic scintillator were a bundle of fibers,
each optically coupled to a single pixel in an imager that may have $> 10^6$
pixels), provided that the flux is predictable to a factor of ${\cal O}(1)$.
If not predictable, as is likely when source temperatures are low because of
the extreme temperature sensitivity of the reaction rates, either too few
neutrons would be detected for statistical significance or too many.  In a
segment in which $\gtrsim 5\,$MeV is deposited it would not be possible to
distinguish a number of 14.1 MeV neutrons from several times that number of
2.45 MeV neutrons.
\subsection{Activation}
%
%
DT neutrons may activate nuclei that cannot be activated by the lower energy
DD neutrons.  A promising candidate, on account of its cross-section,
half-life and detectability of the product, is the
$^{27}$Al(n,$\alpha$)$^{24}$Na reaction with a threshold of 3.25 MeV.
Detection would require rapid processing (the half-life of $^{24}$Na is only
15 hours, while those of most other candidate targets are even shorter) of
large quantities of material.

The cross-section for this reaction at 14.1 MeV is about 0.12 b, while the
total scattering cross section is about 1.75 b and inelastic scattering
($^{27}$Al has low-lying states at 0.844 MeV and 1.014 MeV and many more
states above 2 MeV) has a cross-section of about 0.4 b at 14.1 MeV, and
somewhat greater at energies between 2 and 10 MeV.  In a thick Al slab the
fraction of incident neutrons producing activation is no more than 10\%;
even though energy loss by recoil in elastic scattering is small (roughly
1/27 of the neutron's incident energy), energetic neutrons lose energy
rapidly by inelastic scattering.  A thick slab must have a column density
$> 25\,$g/cm$^{2}$ (thickness $> 10\,$ cm), a mass of hundreds of kg for a 1
sterad activation target at a distance of 1 m from the source.  The total
activation efficiency, allowing for solid angle and the cross-sections, is
then $\lesssim 1\%$.  This massive slab of Al must be dissolved and a
handful of $^{24}$Na nuclei efficiently separated (perhaps by
reprecipitation of the Al and extraction of Na from the supernatant) in a
few hours, and their decay gamma rays efficiently counted.  Although a
conceivable means of detecting yields $> 10^3$ of DT neutrons, and of
measuring yields $> 10^4$--$10^5$ neutrons with useful accuracy, this would
be cumbersome.
\subsection{Inelastic scattering}
\label{inelastic}
DT neutrons (14.1 MeV) may produce prompt $\gamma$-rays by inelastic
scattering to levels that cannot be reached by scattering of the 2.45 MeV
neutrons of DD reactions.  Carbon is a uniquely favorable target because
both its stable isotopes have thresholds for excitation above 2.45 MeV
(for other elements, even oxygen, some isotope has a lower excitation
threshold) and for its ready availability and convenience.  A scintillator
shielded by hydrogenous material would detect the $\gamma$-rays.

\begin{figure}
\begin{center}
\includegraphics[width=4in]{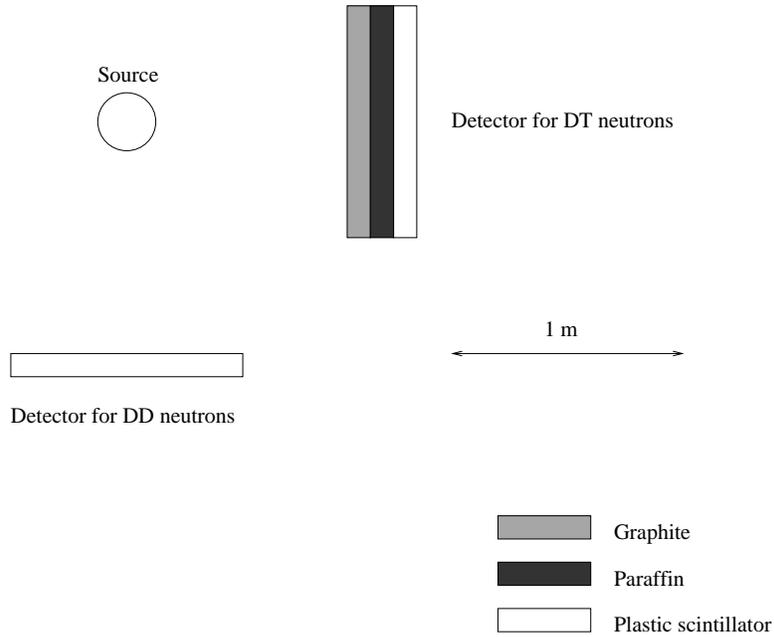}
\end{center}
\caption{Detector system, approximately to scale, to measure and
distinguish DT and DD neutrons from a weak neutron source.}
\label{DTDDdetector}
\end{figure}

A possible design is shown in Fig.~\ref{DTDDdetector}.  The first slab
(closest to the source) is made of graphite.  For a 14.1 MeV neutron the
cross-section $\sigma_{n,\gamma}$ for excitation of the first excited state,
which is followed by emission of a 4.44 MeV $\gamma$-ray, is 0.21 b.  This
process competes with the total cross-section $\sigma_t$ of 1.32 b.  Neutron
transport is complicated, but for the purposes of an analytic estimate we
make the (conservative) approximation that the incident neutron flux is
exponentially attenuated at the rate $\beta \equiv n_C \sigma_t =
0.145$/cm, where for a graphite density of 2.16 gm/cm$^3$ $n_C = 1.1 \times
10^{23}$ cm$^{-3}$.  This ignores excitation by scattered neutrons, whose
energy remains well above the excitation threshold even after ${\cal O}$(10)
elastic scatterings ($\sigma_{n,\gamma}$ is roughly constant in the range
6--14 MeV), although inelastic scattering into the $3\alpha$ channel at
energies above 7.89 MeV, included in $\sigma_t$, reduces the excitation by
scattered neutrons.  

The neutron to gamma-ray conversion rate $\gamma \equiv n_C 
\sigma_{n,\gamma} = 0.23$/cm, and 4.44 MeV gamma rays are attenuated in
graphite at a rate $\alpha \equiv n_C \sigma_{abs} = 0.063$/cm
($\sigma_{abs} = 0.57$ b, essentially the Compton scattering cross-section).
We make a one-stream approximation but allow for the fact that half of the
emitted gamma rays are directed backwards by using $\gamma^\prime = \gamma/
2$, and take the attenuation coefficient for gamma rays at the mean angle to
the normal of an isotropic distribution ($60^\circ$): $\alpha^\prime = 2
\alpha$.  The latter approximation is conservative because attenuation
collimates the gamma ray flux into small angles to the slab normals with an
effective attenuation coefficient close to $\alpha$ rather than
$\alpha^\prime$.  The gamma ray flux at a depth $z$ into the graphite slab
\begin{equation}
f_{\gamma} = {f_{DT}\gamma^\prime \over \alpha^\prime - \beta}
\left[\exp{(-\beta z)} - \exp{(-\alpha^\prime z)}\right],
\end{equation}
where $f_{DT}$ is the incident 14.1 MeV neutron flux.
This is maximized at a depth, corresponding to the optimal slab thickness,
\begin{equation}
z_{opt} = {\ln{(\beta/\alpha^\prime)} \over \beta - \alpha^\prime} = 
7.4\,\mathrm{cm} \approx {1 \over \beta}.
\end{equation}
The emergent 4.44 MeV $\gamma$-ray flux, in this approximation, is
\begin{align}
\begin{split}
f_\gamma &= f_{DT} {\gamma^\prime \over \alpha^\prime - \beta} 
\left[\exp{(-\beta z_{opt})} - \exp{(-\alpha^\prime z_{opt})}\right] \\ &= 
f_{DT} {\gamma^\prime \over \alpha^\prime - \beta} 
\left[\left({\beta \over \alpha^\prime}\right)^{-\beta/(\beta - \alpha^\prime)} -
\left({\beta \over \alpha^\prime}\right)^{-\alpha^\prime/(\beta - \alpha^\prime)}\right] \\ 
&= 0.031 f_{DT}.
\label{flux4.4}
\end{split}
\end{align}
A 1 m$^2$ slab (about 1 sterad at 1 m distance), 7.4 cm thick, of graphite
has a mass of 160 kg.  It need not be high purity, and (unlike an aluminum
activation target) may be reused indefinitely without processing.
\subsection{Filtering}
Graphite has a scattering cross section to 2.45 MeV neutrons of 1.59 b, so
that their unscattered flux is attenuated at a rate $\beta_{DD} = 0.175$/cm
and the flux that emerges unscattered from a slab of thickness $z_{opt}$ is
\begin{equation}
\label{flux2.45}
f_{2.45} = f_{DD} \exp{(-\beta_{DD} z_{opt})} = 0.27 f_{DD}.
\end{equation}
This compares unfavorably with the 4.44 MeV $\gamma$-ray flux 
Eq.~\ref{flux4.4}, and it is necessary to attenuate the lower energy
neutrons.

These can be attenuated by a slab of hydrogenous material such as paraffin
wax or polyethylene, as shown in Fig.~\ref{DTDDdetector}.  Because of the
large neutron scattering cross-section of hydrogen at 2.45 MeV (2.6 b), the
total cross-section of wax per carbon atom is 6.8 b, and at a density of
0.9 g/cm$^{3}$ the unscattered 2.45 MeV neutron flux is attenuated at a rate
of 0.26/cm.  In contrast, the cross-section of wax per carbon atom for 4.44
MeV {\it photons\/} is only 0.76 b, and they are attenuated at a rate of
0.060/cm, where we have again taken a mean angle to the normal of
$60^\circ$.  The scattering cross-section of wax for 14.1 MeV neutrons is
2.66 b per carbon, and the attenuation is 0.10/cm.

A thickness of 10 cm of wax attenuates the unscattered 2.45 MeV neutrons by
a factor of 0.074, but the 4.44 MeV gamma rays by a factor of 0.55.  The
resulting flux ratio, using Eqs.~\ref{flux4.4} and \ref{flux2.45},
\begin{equation}
{f_\gamma \over f_{2.45}} = {0.55 \over 0.074} {0.031 f_{DT} \over 0.27
f_{DD}} =  0.85 {f_{DT} \over f_{DD}}.
\label{fom}
\end{equation}
Each additional 10 cm of paraffin multiplies the coefficient in Eq.~\ref{fom}
by a factor of $0.55/0.074 = 7.4$ while reducing the $\gamma$-ray sensitivity
by only 45\%, so that a thick sandwich detector is effectively only sensitive
to 14.1 MeV neutrons and the gamma-rays they produce.
\subsection{Data Inversion}
DD neutrons may be detected by a simple scintillator detector.  The more
energetic DT neutrons and the gamma rays they produce will also excite the
scintillator, so it produces a weighted sum signal.  Once calibrated, the
fluxes of the two energies of neutrons may be found from the signals in the
two scintillators by inverting the response matrix $\mathbf{R}$.  This
matrix is defined by the relation between the numbers $N_i$ of source
neutrons and the energies $E_i$ deposited in the $i$-th detector:
\begin{equation}
\begin{pmatrix}
E_1 \\
E_2
\end{pmatrix}
=
\begin{pmatrix}
r_{11} & r_{12} \\
r_{21} & r_{22}
\end{pmatrix}
\begin{pmatrix}
N_1 \\
N_2
\end{pmatrix}
,
\end{equation}
where $i = 1$ denotes 14.1 MeV neutrons and the unfiltered scintillator, and
$i = 2$ denotes 2.45 MeV neutrons and the filtered detector.
\section{Quantitative Results}
We use the Monte-Carlo simulation code MCNP6 to calculate the energy
deposited in the scintillators in the detector geometry of
Fig.~\ref{DTDDdetector} when subjected to fluences of 2.45 and 14.1 MeV
neutrons.  The slabs are circular discs 1 m in diameter, and the centers of
the closest surfaces of each detector array are 1 m from a point neutron
source at the origin, so that the front surfaces of each detector subtend a
solid angle 0.663 sterad at the neutron source, or 0.0528 of the sphere.
The calculation includes all relevant neutron, electron and photon
processes.

We define a figure of merit of the detector system:
\begin{equation}
\label{fomeq}
\mathrm{FOM} \equiv {r_{21} \over r_{22}} {r_{12} \over r_{11}}.
\end{equation}
The first factor is the ratio of energy deposited by 2.45 MeV neutrons
into the unfiltered detector to that deposited into the detector with
graphite converter and filter, and the second factor is the ratio of 
energy deposited by 14.1 MeV neutrons into the detector with converter
and filter to that deposited into the unfiltered detector.  Ideally, these
ratios would be infinite, so that the unfiltered detector would only detect
2.45 MeV neutrons and the detector with converter and filter would only
detect 14.1 MeV neutrons.  Conversely, if the two detectors had the same
response to neutrons of each energy the factors would be unity, $\mathbf{R}$
would be singular, $\mathrm{FOM} = 1$ and it would be impossible to
determine the separate production rates of 2.45 MeV and 14.1 MeV neutrons.

The accuracy with which the neutron sources can be inferred from the data is
limited by statistical uncertainty in the energy deposited in the
scintillators because of the finite number of deposition events.  The system
maximizes $\mathrm{FOM}$ by heavily filtering the flux in detector 2,
reducing the energy deposited and increasing its statistical uncertainty.
The optimum choice of detector parameters requires a tradeoff among these
criteria and other factors such as cost.

\begin{table}[h!]
\begin{center}
\begin{tabular}{|r|c|c|c|c|c|}
\hline
& graphite & paraffin & scintillator & $\mathrm{FOM}$ &
$\sigma_2 \sqrt{N_1/10^4}$ \\
\hline
1 & 10 & 10 & 10 & 2.25 & 0.089\\
2 & \phantom{0}5 & \phantom{0}5 & 10 & 1.69 & 0.063 \\
3 & \phantom{0}2 & \phantom{0}2 & 10 & 1.25 & 0.054 \\
4 & \phantom{0}1 & \phantom{0}1 & 10 & 1.11 & 0.047 \\
5 & 15 & 15 & 10 & 2.53 & 0.123 \\
6 & 20 & 20 & 10 & 2.57 & 0.168 \\
7 & 20 & 10 & 10 & 2.28 & 0.123 \\
8 & 10 & 15 & 10 & 2.53 & 0.104 \\
9 & 10 & 20 & 10 & 2.68 & 0.120 \\
10 & 10 & 25 & 10 & 2.72 & 0.142 \\
11 & 10 & 10 & \phantom{0}5 & 2.21 & 0.092 \\
12 & 10 & 10 & 20 & 2.38 & 0.092 \\
13 & \phantom{0}0 & 10 & 10 & 1.89 & 0.047 \\
\hline
\end{tabular}
\end{center}
\caption{\label{detectortable}Detector performance.  The first three columns
indicate the thicknesses (in cm) of the graphite neutron to gamma-ray
converter and paraffin neutron filter for detector 2 and (for both
detectors) the thickness of the plastic scintillator.  $\mathrm{FOM}$ is the
system figure of merit defined in Eq.~\ref{fomeq}.  The final column shows
the fractional statistical uncertainty in the energy deposited in the 
filtered detector 2 by 14.1 MeV neutrons and their products, the critical
uncertainty in determining $N_1/N_2$; $\sigma_2 \propto N_1^{-1/2}$.  The
calculational uncertainties in the $\mathrm{FOM}$ are about $\pm 0.01$.}
\end{table}

Table~\ref{detectortable} presents the results of model calculations. 
These results show that $\mathrm{FOM}$ increases only slowly with
converter and filter thicknesses beyond 10 cm, but that increased thickness
significantly increases the statistical uncertainty, so that values near 10
cm appear to be near optimal.  Increasing the scintillator thickness in the
baseline case (line 1) from 10 cm to 20 cm (line 12) increases the energy
deposited in the filtered detector 2 by a source of 14.1 MeV neutrons by
60\%, but does not reduce its statistical uncertainty.

The Monte-Carlo results also show that typically about 70\% of the energy
deposition in the scintillator of detector 2 is by neutrons.  The graphite
turns a portion of the energy of 14.1 MeV neutrons into gamma rays, but it
and the paraffin filter also function as a neutron energy discriminator,
preferentially attenuating lower energy neutrons.  Comparison of the
baseline line 1 with line 13 in which the graphite is omitted shows that the
graphite significantly increases the $\mathrm{FOM}$, but that the paraffin
filter alone discriminates between 14.1 MeV and 2.45 MeV neutrons.

Statistical uncertainties may be estimated from entries in the last column
that show the fractional standard deviation $\sigma_2$ of $E_2$ 
(contributed by the 14.1 MeV neutron source that typically accounts for
80\% of $E_2$).  The Monte Carlo simulations used $N_1 = N_2 = 10^7$ to
minimize simulation (as opposed to experimental) statistical uncertainty.
The statistical uncertainty $\sigma_1$ of $E_1$ in the unfiltered detector
is generally smaller because several times as much energy is deposited in
it, so $\sigma_2$ is a fair estimate of the fractional uncertainty in $N_1$.
Because the 2.45 MeV neutrons principally contribute to $E_1$ (with its
smaller uncertainty) the fractional uncertainty in $N_2$ will generally be
smaller than that in $N_1$.

Although $\sigma_2$ is only one contribution to uncertainty in the desired
ratio $N_1/N_2$, provided $\mathrm{FOM} \gtrsim 2$ it is the dominant source
of uncertainty.  Hence, if $N_1 \gtrsim 10^4$ and $N_2 \gtrsim 10^4$ the
ratio $N_1/N_2$ may, with good choice of detector parameters, be determined
to a $1 \sigma$ accuracy of better than 20\%.  At tempertures of a few tens
of eV this corresponds to a comparable fractional uncertainty in
temperature (the logarithmic slope of the solid line in
Fig.~\ref{reactratio} is close to unity).  Because of the extreme
temperature sensitivity ($\propto T^{15}$; Fig.~\ref{reactratio}) of both
reaction rates in this temperature range, this implies a roughly ten-fold
uncertainty in $\int_{T \ge 0.93T_{max}} n^2\,dV$, where $n$ is the particle
density and the integral is taken over the region in which the temperature
is within 7\% of its maximum value.  In this region the reaction rate is
within a factor of three of its maximum, and the integral is a fair
approximation to $\int \langle \sigma v \rangle(T) n^2\,dV / \langle \sigma
v \rangle(T_{max})$, a single parameter description of the reaction region.

I thank Carl Hagelberg for essential discussions.  The Los Alamos National
Laboratory is operated by Los Alamos National Security, LLC for the U.~S.
Department of Energy under Contract No.~DE-AC52-06NA25396.

\end{document}